\documentclass[twocolumn,aps,prc,showpacs,preprintnumbers]{revtex4}%
\usepackage{amssymb}
\usepackage{amsmath}
\usepackage{graphicx}
\usepackage{dcolumn}
\usepackage{bm}
\usepackage{amsfonts}%
\begin{document}
\title{Electro-disintegration following beta-decay}
\author{C.A Bertulani$^{(1,2)}$}
\email{bertulanica@ornl.gov}
\affiliation{{$^{1}$Department of
Physics and Astronomy University of Tennessee Knoxville, Tennessee
37996, USA \\
$^{2}$Physics Division Oak Ridge National Laboratory P.O. Box 2008
Oak Ridge, Tennessee 37831, USA}}

\begin{abstract}
I show that the disintegration of weakly-bound nuclei and the
ionization of weakly-bound atomic electrons due to their interaction
with leptons from beta decay is a negligible effect.

\end{abstract}
\maketitle
\date{\today}

The disintegration of weakly bound nuclei with small neutron
separation energy in stars can impose limits to the stellar scenario
where these nuclei exist. Beta-decay already sets stringent limits
on the existence of nuclei very far from the line of stability (see,
e.g. \cite{Ri02}). Here I discuss an additional effect, namely the
restrictions imposed by final state interactions of the
beta-particle with the daughter nucleus. Electrons observed in
beta-decay can have enough kinetic energy to induce the dissociation
of the daughter nucleus with small separation energy.  If this
process is proven to be relevant, it would lead to the existence of
voids in the elemental abundance close to the drip-line.

The basic assumptions adopted here are that the excitation
(dissociation) of a nucleus following beta-decay is sequential and
that it can be described as a two-step process, so that the
transition rate is given by
\[
W_{i\rightarrow m\rightarrow f}=W_{i\rightarrow m}^{\left(  \beta\right)
}\cdot P_{m\rightarrow f}^{(e)}
\]
where $W_{i\rightarrow m}^{\left(  \beta\right)  }$ is the usual
beta-decay transition rate from an initial nuclear state $i$ to an
intermediary state $m $, and $P_{m\rightarrow f}^{(e)}$\ is the
probability for the nuclear excitation from $m$ to a final state $f$
by the interaction of the nucleus with the outgoing electron
(positron).

The beta-particle is described by a spherically symmetric outgoing
wave, that favors monopole transitions in the daughter nucleus. We
neglect retardation and assume that the electron (positron) energy
is much larger than the excitation energy. The outgoing electron
wave will generate a time-dependent monopole wake field
whose interaction with the nucleus has the usual form $V_{e}=e_{eff}%
/r$, where $e_{eff}$ is the effective charge for the transition. The
effective charge arises due to the modification of the charge radius
of the nucleus after nucleon emission. An accurate value of the
effective charge depends strongly on the nuclear properties
\cite{Ta72}. For simplicity, I will assume $e_{eff}\sim e$.

Because of the assumed spherical symmetry, the Coulomb field of the
electron (positron) only exists outside the outgoing electron
wavefront. Therefore, in first-order time-dependent perturbation
theory, the excitation amplitude $A_{m\rightarrow
f}^{(e)}$ is given by%
\begin{align}
A_{m\rightarrow f}^{(e)}  & =\frac{1}{i\hbar}e^{2}\int dt\exp\left[  i\left(
E_{f}-E_{m}\right)  t/\hbar)\right] \nonumber\\
& \int_{r>r_{e}(t)}d^{3}r\ \frac{1}{r}\Psi_{f}^{\ast}\left(  \mathbf{r}%
\right)  \Psi_{m}\left(  \mathbf{r}\right)  ,\label{afi}%
\end{align}
where we set up the spin angular part of the matrix element equal to
1. $\Psi_{j}\left(  \mathbf{r}\right)  $\ denotes the nuclear
wavefunction, $E_{j} $ the nuclear energy of state $j$, $r_{e}$ is
the electron and $r$ the internal nuclear coordinate.

We use a simplified nuclear model for the nuclear wavefunction $\Psi\left(
\mathbf{r}\right)  $ which captures the essence of the process. The
wavefunction for the state $m$ is taken as an $s$-wave Hulth\'{e}n $\ $wave
function \cite{HS57}%
\begin{equation}
\Psi_{m}\left(  \mathbf{r}\right)  =N\frac{u_{m}(r)}{r}=N\frac{\left(
e^{-\alpha r}-e^{-\beta r}\right)  }{r}.\label{hulthen}%
\end{equation}
The term $e^{-\beta r}$ modifies the asymptotic form $e^{-\alpha r}$
at small distances in such a way that $u_{m}(0)=0$, and more
specifically $u_{m}\sim r$, as is reasonable for $s$ waves.
Moreover, the parameter $\alpha$ is given in terms of the separation
energy of the nucleon from the nucleus by the equation
$S_{n}=\hbar^{2}\alpha^{2}/2m_{n}$, where $m_{n}$ is the reduced
nucleon-nucleus mass and $\beta$ can be determined from the
effective range
parameter, $r_0$, as approximately \cite{HS57,Adl70}%
\begin{equation}
\beta=\frac{3-\alpha r_{0}+\left(  \alpha^{2}r_{0}^{2}-10\alpha r_{0}%
+9\right)  ^{1/2}}{2r_{0}},\label{param}%
\end{equation}
and in general $\beta\gg1$. Similarly, the normalization constant
can be expressed in terms of the effective range as
$N^{2}=\alpha\left[ 2\pi(1-\alpha r_{0})\right]  ^{-1}$.\ In the
following numerical calculations we will use $r_{0}=3$ fm, a typical
value for nuclear systems.

The final wavefunction is an outgoing spherical wave,
\begin{equation}
\Psi_{f}\left(  \mathbf{r}\right)  =\frac{1}{2ikr}\exp\left(  ikr\right)
\label{wfinal}%
\end{equation}
where $k$ is related to the relative kinetic energy of the final state by
$\varepsilon=\hbar^{2}k^{2}/2m_{n}$.%

\begin{figure}
[ptb]
\begin{center}
\includegraphics[
height=2.8089in,
width=2.8902in
]%
{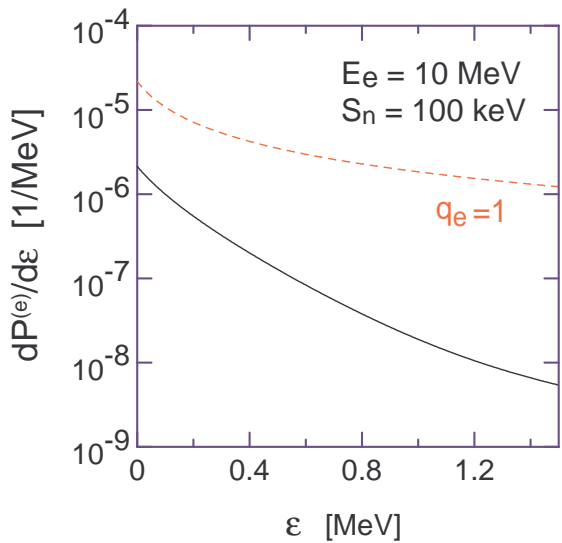}%
\caption{(Color online) Energy spectrum of continuum states produced
by electro-dissociation following a beta-decay with electron
(positron) energy $E_{e}=10$ MeV. The initial state is bound by 100
keV. The dashed curve is obtained with the approximation of
eq. \ref{appro}. }%
\label{f1}%
\end{center}
\end{figure}

If $v_{e}$ denotes the electron velocity, assumed to remain
undisturbed by the energy transfer to the excitation, the time
dependence of the electron position is $r_{e}=v_{e}t$. The first
integral in eq. \ref{afi} can be
expressed in terms of the exponential integral function, $\operatorname{Ei}%
\left(  x\right)  $, as%
\begin{align}
A_{m\rightarrow f}^{(e)}\left(  E_{e},S_{n},\epsilon\right)   & =\frac{2\pi
e^{2}N}{\hbar kv_{e}}\int_{0}^{\infty}dr\ q_{e}\left(  \lambdabar
_{e},r\right)  \ \exp\left(  i\frac{\omega r}{v_{e}}\right) \nonumber\\
& \left[  \operatorname{Ei}\left(  -ar\right)  -\operatorname{Ei}\left(
-br\right)  \right]  ,\label{amf}%
\end{align}
where $E_{e}$ is the electron (positron) energy, $a=\alpha+ik$ and\
$b=\beta+ik$ and we use the short notation $\omega=\left(
E_{f}-E_{m}\right)  /\hbar,$ such that
$\hbar\omega=S_{n}+\varepsilon$. Note that we have introduced an
electron-charge distribution $\ q\left(  \lambdabar_{e},r\right)  $
which has the following meaning. When the electron (positron) is
produced in beta-decay its charge is homogeneously distributed
within a sphere of the size of its Compton wavelength
$\lambdabar_{e}=\hbar/\gamma m_{e}c$, where $\gamma=\left(
1-v_{e}^{2}/c^{2}\right)^{-1/2}  .$ This is based on the uncertainty
principle, which introduces a smearing out of the electron
coordinate within a region equal to its wavelength. This condition
implies that
\begin{equation}
q_{e}\left(  \lambdabar_{e},r\right)  =\left\{
\begin{array}
[c]{c}%
r^{3}/\lambdabar_{e}^{3},\text{ \ \ \ \ for \ \ }r<\lambdabar_{e}\\
1\text{, \ \ \ \ for \ \ }r\geqslant\lambdabar_{e}%
\end{array}
\right.  .\label{qe}%
\end{equation}

If $q_{e}=1$ is used, the integral in eq. \ref{amf} can be performed
analytically. One gets%
\begin{align}
& \frac{v_{e}}{2\omega}\left\{  2\arctan\left(  \frac{\omega}{v_{e}a}\right)
-2\arctan\left(  \frac{\omega}{v_{e}b}\right)  \right. \nonumber\\
& \left.  -i\left[  \ln\left(  1+\frac{\omega^{2}}{v_{e}^{2}a^{2}}\right)
-\ln\left(  1+\frac{\omega^{2}}{v_{e}^{2}b^{2}}\right)  \right]  \right\}
.\label{appro}%
\end{align}

Finally, the dissociation probability is given by%
\begin{align}
P_{m\rightarrow f}^{(e)}\left(  E_{e},S_{n}\right)   &  =\int d\varepsilon
\rho\left(  \varepsilon\right)  \left\vert A_{m\rightarrow f}^{(e)}\right\vert
^{2}\nonumber\\
&  =\frac{\left(  2m_{n}\right)  ^{3/2}}{(2\pi\hbar)^{3}}\int_{0}^{\infty
}\left\vert A\left(  E_{e},S,\varepsilon\right)  \right\vert ^{2}%
\sqrt{\varepsilon}d\varepsilon,\label{prob}%
\end{align}
where $\rho\left(  \varepsilon\right)  =\left(  2m_{n}\right)  ^{3/2}%
\sqrt{\varepsilon}d\varepsilon/(2\pi\hbar)^{3}$ is the density of final states
of the nucleon-nucleus system.

Figure \ref{f1} shows the energy spectrum, $dP^{(e)}/d\varepsilon$,
of continuum states produced by electro-dissociation following a
beta-decay with electron (positron) energy $E_{e}=10$ MeV. The
initial state is bound by 100 keV. The dashed curve is obtained with
the approximation of eq. \ref{appro}. One sees that, as expected,
neglecting the wave character of the electron (i.e., using eq.
\ref{appro}) leads to a large overestimation of the excitation
probabilities. Using the value of $q_{e}$ as given by eq. \ref{qe}
leads to a steeper decrease of states with larger energy. Obviously,
for too large excitation energies of the nucleus the total energy is
not conserved and the formalism described above is not appropriate.

\begin{center}%
\begin{tabular}
[c]{cc}\hline $S_{n}\ $[keV] & $P^{(e)}$\\\hline 10 &
$9.3\times10^{-7}$\\ 70 & $4.8\times10^{-7}$\\ 160 &
$1.9\times10^{-7}$\\ 310 & $7.3\times10^{-8}$\\\hline
\end{tabular}

\end{center}

{\small Table 1: Dissociation probability of a loosely-bound nucleus as a
function of the neutron separation energy $S_{n}$ in keV for an electron with
energy $E_{e}=10$ MeV.}

\bigskip

Table 1 shows the dissociation probability of a loosely-bound
nucleus as a function of the neutron separation energy $S_{n}$ in
keV for an electron (positron) with energy $E_{e}=10$ MeV. The
probabilities are very small, even for 10 keV separation energy.
This rules out nuclear dissociation following beta-decay as a
relevant effect in beta-decay processes close to the drip line. A
full quantum mechanical calculation will not change this conclusion
as the main ingredients of the effect have been taken into account
above. Also, for charged particle (e.g., emission of a proton) this
effect is further suppressed due to the Coulomb barrier.

Na\"{\i}vely, this calculation can be used to estimate the
probability that the beta-particle ionizes the atom by ejecting one
of its outer electrons. One can use the equations above and just
replace the nucleon mass by the electron mass (using $r_0=0$). While
the Hulth\'{e}n wavefunction, eq. \ref{hulthen}, is a good
approximation for a loosely bound electron, the scattering wave, eq.
\ref{wfinal}, is obviously wrong as it does not account for the
(screened) charge of the residual atom. An estimate of the Coulomb
effect follows by adding a Coulomb phase, $(e^{2}/\hbar
v_{e})\ln\left( 2k_{e}r\right) $, to the exponent in eq.
\ref{wfinal}. It has been checked numerically that this changes the
results by only few percent. Moreover, an exact treatment of Coulomb
distortion tends to decrease the magnitude of the ionization
probabilities in projectile impact processes \cite{BS77}.

Results for atomic ionization following beta decay are shown in
Table 2 as a function of the beta-particle energy assuming a loosely
bound electron with separation energy of $S_{e}=1$ eV. One sees, as
expected, that the ionization probability decreases with the
beta-decay electron energy. The obvious reason is the increase of
the wavelength mismatch between the emitted electron and that of the
beta-particle as the energy of the later increases. The ionization
probability remains small even when the beta-particle has small
energy.

\begin{center}%
\begin{tabular}
[c]{cc}\hline $E_{e}$ & $P^{(e)}$\\\hline 10 eV &
$4.7\times10^{-8}$\\ 50 keV & $6.3\times10^{-10}$\\ 1 MeV &
$1.09\times10^{-10}$\\ 5 MeV & $1.97\times10^{-11}$\\\hline
\end{tabular}
\end{center}
{\small Table 2: Ionization probability of a loosely-bound atom
($S_{e}=1$ eV) as a function of the beta-particle energy $E_{e}$. }

\bigskip\bigskip\bigskip\bigskip\bigskip\bigskip\bigskip\bigskip\bigskip\bigskip\bigskip

We conclude that the excitation, or dissociation, of nuclei as well
as the atomic ionization by the electron (or positron) emitted in
beta-decay processes are negligible effects. A calculation using
Feynman diagram techniques with proper account of relativistic
effects and energy conservation is very unlikely to change these
conclusions. The same line of thought applies to the consideration
of higher multipole interactions.

\bigskip\bigskip\bigskip\bigskip\bigskip\bigskip

{\bf Acknowledgements}

\medskip

I thank Alex Brown for bringing this problem to my attention and to
Akram Mukhamezhanov for useful discussions. This research was
supported by the U.S. Department of Energy under contract No.
DE-AC05-00OR22725 (Oak Ridge National Laboratory) with UT-Battelle,
LLC., and by DE-FC02-07ER41457 with the University of Washington
(UNEDF, SciDAC-2).

\end{document}